\documentclass{sig-alternate}
\usepackage{graphicx}
\begin{document}


\title{Mining User Comment Activity for Detecting Forum Spammers in YouTube}
\numberofauthors{1} 
\author{\alignauthor Ashish Sureka \\
\affaddr{Indraprastha Institute of Information Technology, Delhi (IIIT-D)} \\ \affaddr{New Delhi, India} \\ 
\email{ashish@iiitd.ac.in}
} 
\date{8 March 2011}
\maketitle
\begin{abstract}
Research shows that comment spamming (comments which are unsolicited, unrelated, abusive, hateful, commercial advertisements etc) in online discussion forums has become a common phenomenon in Web 2.0 applications and there is a strong need to counter or combat comment spamming.  We present a method to automatically detect comment spammer in YouTube (largest and a popular video sharing website) forums. The proposed technique is based on mining comment activity log of a user and extracting patterns (such as time interval between subsequent comments, presence of exactly same comment across multiple unrelated videos) indicating spam behavior. We perform empirical analysis on data crawled from YouTube and demonstrate that the proposed method is effective for the task of comment spammer detection.
\end{abstract}

\category{H.3.3}{Information Search and Retrieval}[Information filtering]

\terms{Experimentation, Measurement}

\section{Research Motivation and Aim}
Spam in domains such as emails, web-pages, blogs, social networking websites, online discussion forums, wikis and video sharing websites is prevalent and naturally has several negative impacts such as undesirable consumption of computing resources, lowering the reputation or value of the targeted legitimate web application, impacting search engine rankings, overwhelming moderators and administrators, and obstructs and misleads genuine usage of legitimate users and community \cite{Benevenuto2010}\cite{Hayati2009}\cite{Hayati2010}\cite{Heymann2007}. Previous studies show that \textit{comment spam} in \textit{online discussion forums} (the focus of this paper) is prevalent and techniques to counter such type of spam have attracted several researchers' attention \cite{Bhattarai2009}\cite{Yin2009}\cite{Dhinakaran2009}\cite{Yuan2007}\cite{Youngsang2011}. Several \textit{content-based methods} have been proposed to automatically identify spam comments. Content-based methods analyze the text of the post or message (such as checking the presence of pre-defined terms or links) in a forum and infer the likelihood of a message being spam or legitimate. 
\begin{figure}[ht]
\centering
\includegraphics[scale=0.36]{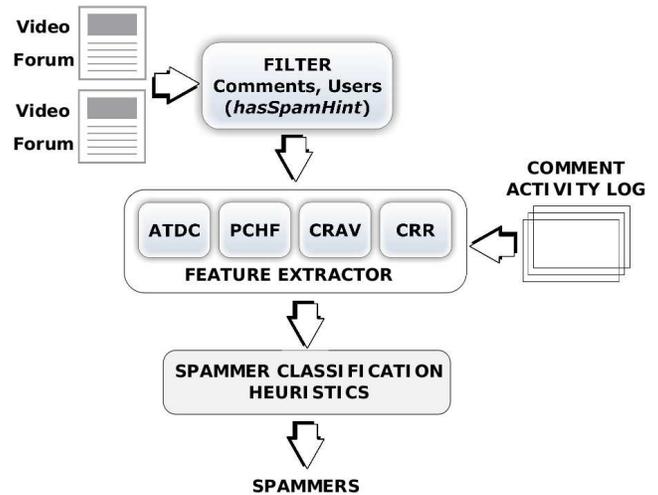}
\caption{High-Level system architecture and processing pipeline for identifying comment spammers in YouTube. ATDC: Average Time Difference between Comments, PCHF: Percentage of comments having hasSpamHint flag, CRAV: Comment repeatability across videos, CRR: Comment repetition and redundancy.}\label{fig:architecture}
\end{figure}
\begin{figure*}[ht]
\begin{minipage}[b]{0.5\linewidth}
\centering
\includegraphics[scale=0.3, angle=270]{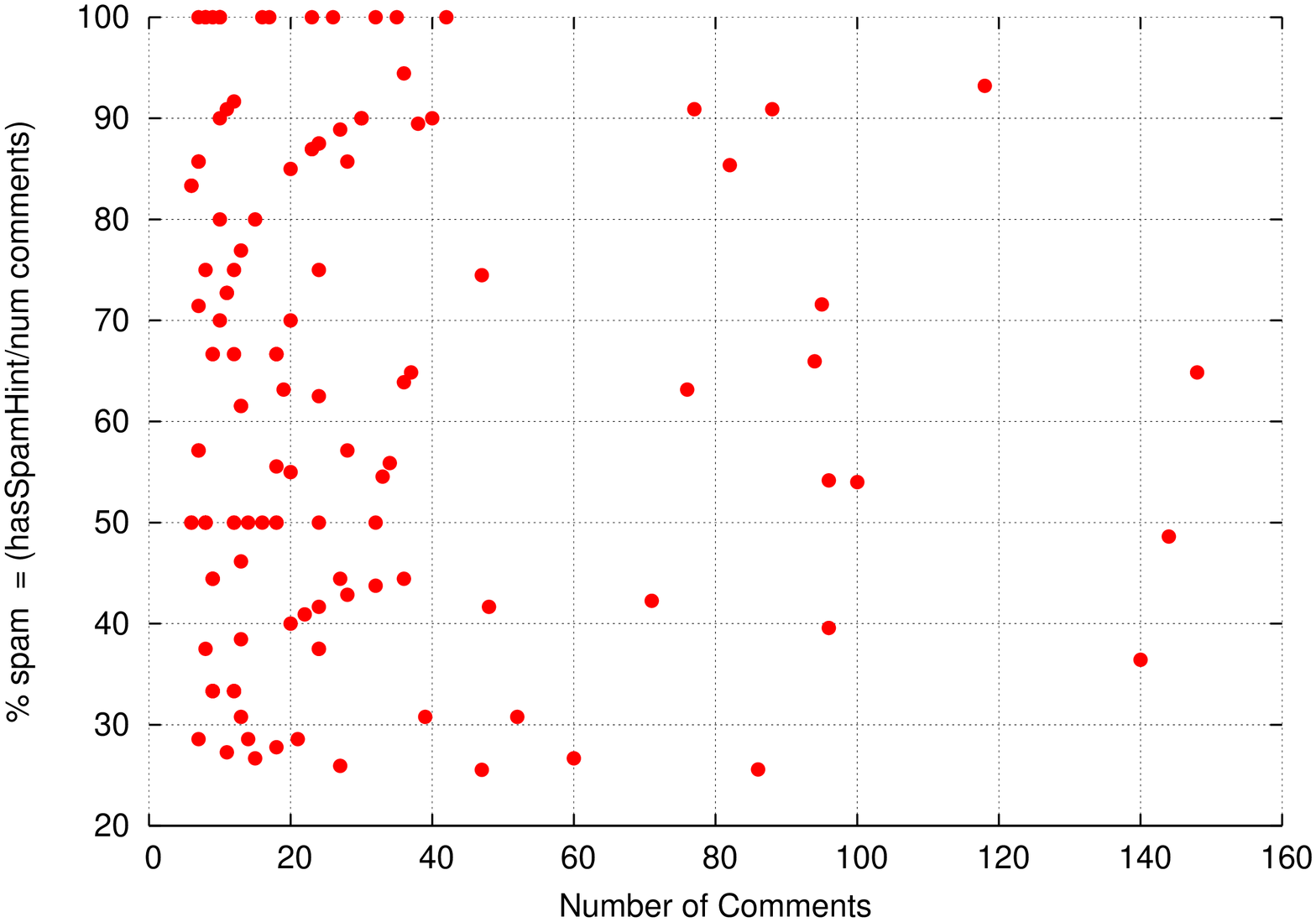}
\caption{Plot of users in the evaluation dataset across two dimensions:  spam percentage and number of comments.}\label{fig:spam}
\end{minipage}
\hspace{0.5cm} 
\begin{minipage}[b]{0.5\linewidth}
\centering
\includegraphics[scale=0.3, angle=270]{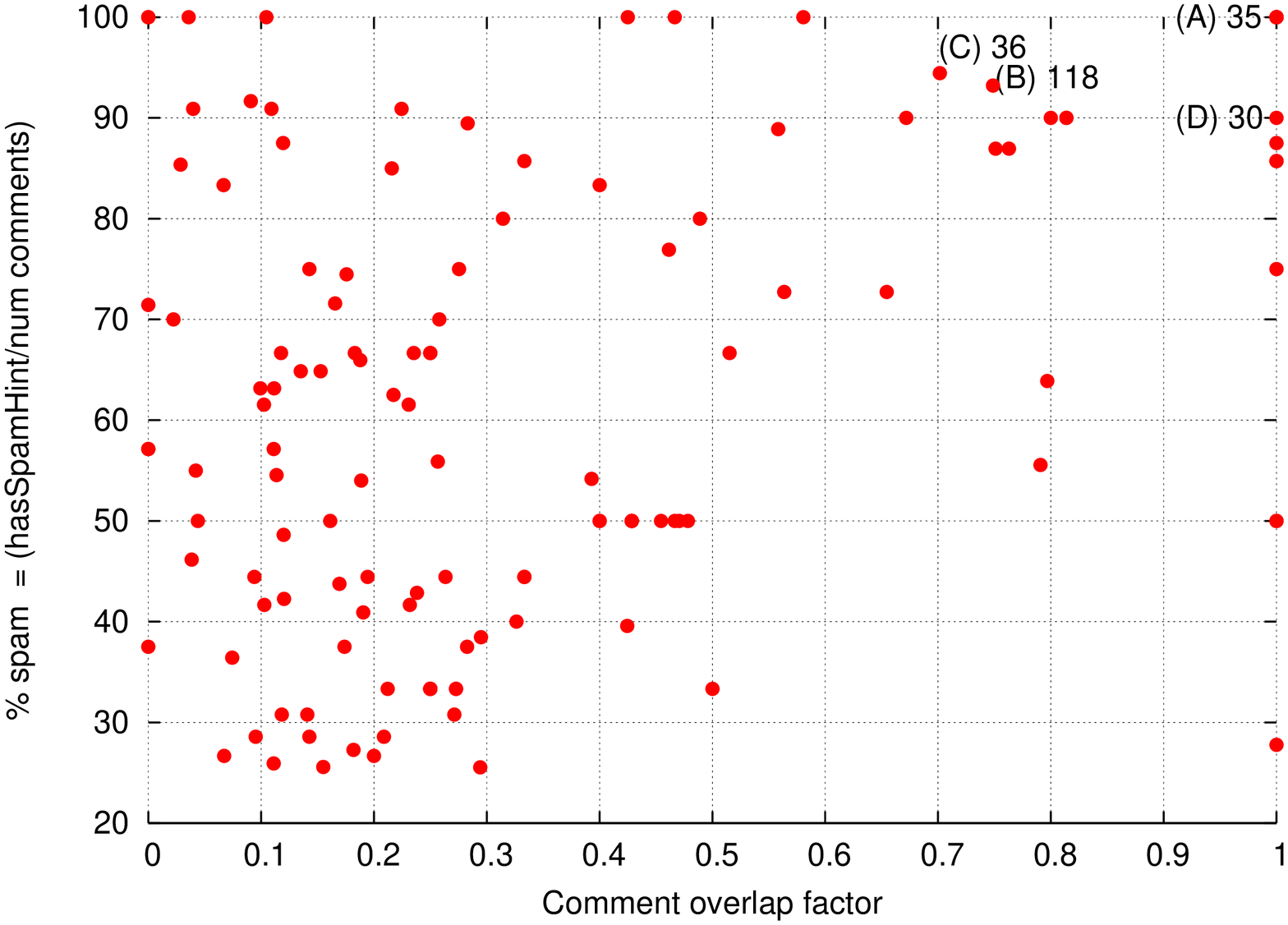}
\caption{Plot of users in the evaluation dataset across two dimensions:  spam percentage and comment repetition and redundancy (CRR).}\label{fig:percspam}
\end{minipage}
\end{figure*}
While content-based methods have shown encouraging results, they are not perfect and there is a strong need to augment or complement the capabilities of existing anti-spam content-based methods to counter the spam problem. Based on our analysis of related body of work and literature, the research area of analyzing the \textit{commenting behavior} or \textit{activity} of a user to identify \textit{spammers} (a user classification task and not a post or message classification task) is an area which is relatively unexplored in contrast to content-based methods. We \textit{hypothesize} and believe that examining the commenting activity (\textit{usage analysis and characterization}) of a user can play a role in identifying spammers.  The \textit{broad research objective} of the work presented in this paper is to investigate the application of \textit{usage-based features} derived from a user's comment activity (by analyzing a \textit{log of recent comments} with associated metadata) to identify comment spammers. The \textit{specific research aim} of the work presented in this paper is to investigate techniques for mining usage-based discriminatory patterns and markers to identify comment spammers in \textit{YouTube} forums (a very popular and largest video sharing website on Internet). 
\section{Research Contributions}
Heymann et al. present a survey of approaches for fighting spam on social websites \cite{Heymann2007}. Hayati presents an evaluation and analysis of Web 2.0 anti-spam methods \cite{Hayati2009}. Benevenuto et al. provide a general overview of pollution in video sharing systems (evidence of pollution, types of pollution, affect on the system and control strategies) such as YouTube \cite{Benevenuto2010}. Yo-Sub Han et al. present an algorithm to evaluate the reputation of a user in YouTube by mining the user's social activity and interactions (such as subscriptions and uploaded contents) \cite{Han2009}. Benevenuto et al. introduce a technique to detect video spammers in YouTube \cite{Benevenuto2008}. The similarity between the study by Yo-Sub Han et al. and Benevenuto et al. (closely related work) and this paper is that the aim of the work is to perform a user classification (particularly automatic user reputation determination, video spammer identification, and comment spammer detection) task in YouTube. The main difference is that in this paper we explore certain commenting activity and attributes (novel in context to current solutions) of a user to detect the likelihood of the user as forum spammer.
In context to closely related work, this paper makes the following novel and \textit{unique contributions}. This paper presents the \textit{first} study (on YouTube) of mining the \textit{recent activity log} of a user to extract \textit{usage-based features} (particularly prevalence of high comment repeatability, presence of exactly same comment across videos, presence of ultra low time difference between comments, presence of a large number of spam tags by the community or moderators) to identify \textit{spammers}. This paper presents an empirical study on dataset crawled from YouTube and demonstrates that the proposed usage-based and behavioral features can be used or exploited as markers for the task of automatically detecting comment spammers in YouTube forums. The paper offers fresh perspective and insights on the characteristics and properties of comment spammers on YouTube. 

\section{Solution Approach}
Figure \ref{fig:architecture} presents a high level solution framework and key components of the proposed systems. YouTube discussion forums (threaded discussions in response to an uploaded video) have a feature in which comments are marked as \textit{hasSpamHint}. We carefully observed (based on manual and visual inspection) several forums of several videos across various categories and notice many comments correctly tagged as hasSpamHint (wherein the comments are still visible). However, we also notice many (significant percentage) spam comments which are not tagged as hasSpamHint (perhaps due to practical infeasibility of manually analyzing very large volumes of comments by administrators). Furthermore, the tagging of hasSpamHint is performed at the comment-level and not at the \textit{user-level\textit{}}. 

The proposed approach consists of first retrieving comments marked with hasSpamHint for a given video.  We then extract userids behind the spam comments. YouTube APIs\footnote{http://code.google.com/apis/youtube/overview.html} provide functions to retrieve the recent commenting activity (a log of comments and the associated metadata) of a given user. As shown in Figure \ref{fig:architecture}, we extract several comment attributes from the discussion-forum usage-log:  text of the comment, timestamp, VideoID of video commented-on and the value of the binary variable hasSpamHint. The next step consists of computing the values of variables indicating the spam intention of user (as spammer). We define four indicators and describe our intuition (and design justification) behind the proposed indicators. The value of the following four indicators (heuristics) is then used to score a give user as comment spammer. 
\subsection{ATDC}Average Time Difference between Comments (ATDC): We extract all the recent comments (the number of comments that can retrieved is limited by YouTube API) by a user and compute the time differences between all the comments (comparing each comment with every other comment in the log).  We compute the average time difference and record the value. We hypothesize that a low value of ATDC signals spam. Our conjuncture is based on the observation that spammers often employ automated scripts or spam robots for posting comment as a result of which the time difference between subsequent comments is so low that it is not manually feasible.  We confirm the presence of the phenomenon (in the evaluation dataset and also based on our manual inspection of several YouTube forums) wherein the time interval between sequences of comment is less than few seconds.  
\subsection{PCHF}Percentage of Comments with hasSpamHint Flag (PCHF): We compute the percentage of comments marked as hasSpamHint. We hypothesize that a significant percentage of comments by a user marked as hasSpamHint can be used as a signal for classifying the user as spammer. We confirm the prevalence of this phenomenon in YouTube user comment logs. We notice several users who are spammers (validated based on manual inspection) exhibit a high PCHF value. 
\subsection{CRAV}Comment Repeatability Across Videos (CRAV): We observe a phenomenon (which is exploited as an attribute and heuristic for spammer categorization task) wherein a user posts exactly same comment across discussion forum accompanying several different videos. A visual inspection of this phenomenon clearly shows that users posting same message across several videos is a case of content promotion and is a reliable comment activity marker for identifying spammers.  A high variability in terms of videoids and a high similarity of comments posted by a user is employed as an indicator in the proposed solution. 
\subsection{CRR}Comment Repetition and Redundancy (CRR): We observe the presence of a pattern wherein a user simply repeats and posts the same message on the same video (sometimes within a small time interval and sometimes reasonable spread across the time dimension but still the same message). We hypothesize that a high value of CRR signals spam. We compare the text of every comment with the text of all other comments in the log of recent comments posted by a user (1 for an exact match and 0 for a non-match) and compute the average CRR value. 
\begin{figure*}[ht]
\begin{minipage}[b]{0.5\linewidth}
\centering
\includegraphics[scale=0.3, angle=270]{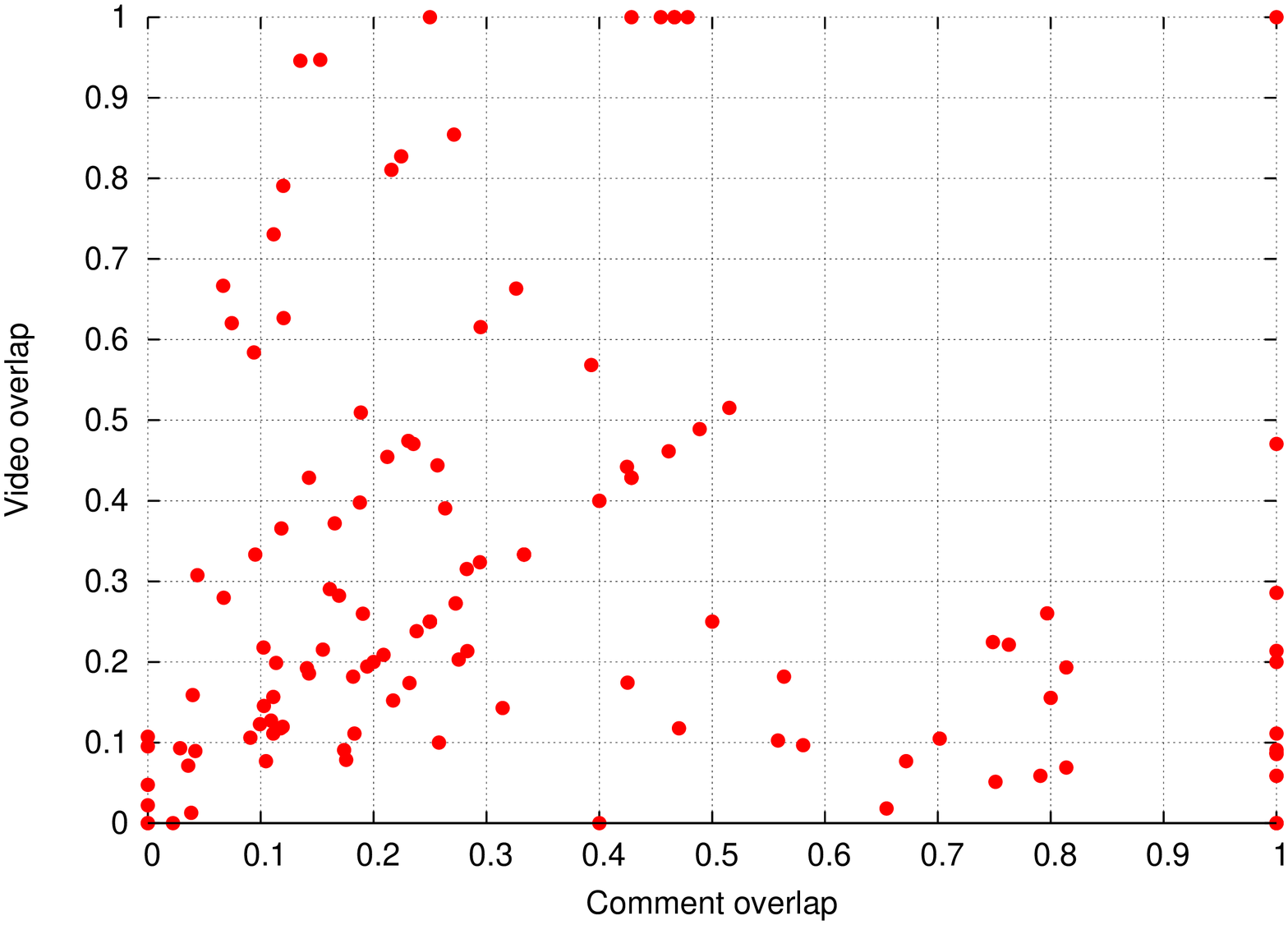}
\caption{Plot of users in the evaluation dataset across two dimensions:  video overlap and comment repetition and redundancy (CRR).}\label{fig:commvideo}
\end{minipage}
\hspace{0.5cm} 
\begin{minipage}[b]{0.5\linewidth}
\centering
\includegraphics[scale=0.3, angle=270]{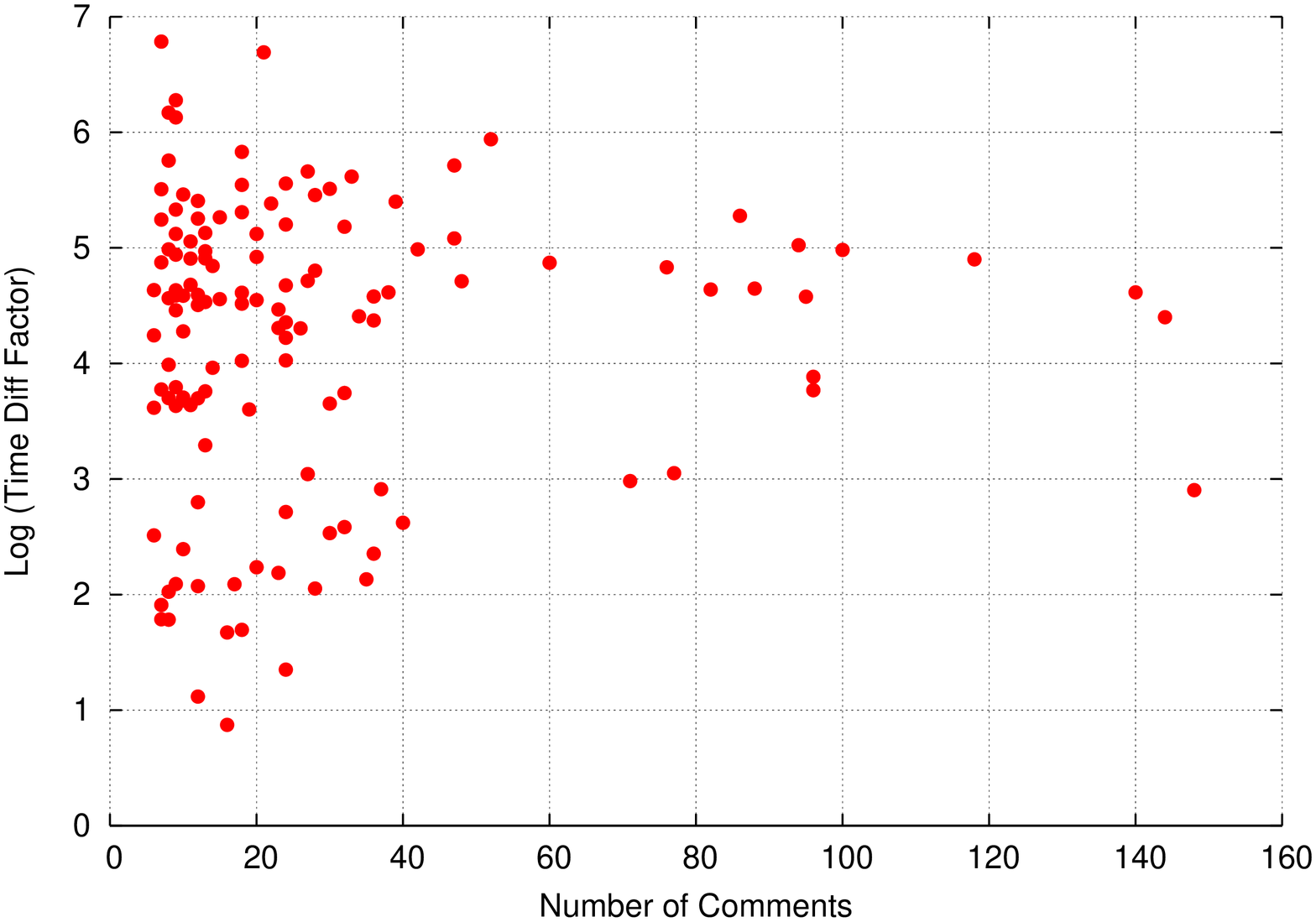}
\caption{Plot of users in the evaluation dataset across two dimensions:  log of average time difference between comments (ATDC) and number of comments.}\label{fig:timediff}
\end{minipage}
\end{figure*}
\section{Empirical Analysis}
\begin{figure}[ht]
\centering
\includegraphics[scale=0.30, angle=270]{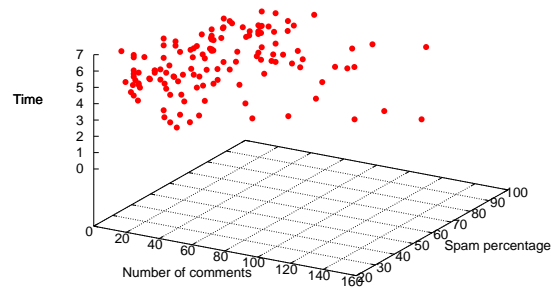}
\caption{Plot of users in the evaluation dataset across three dimensions:  log of average time difference between comments (ATDC),  number of comments and spam percentage.}\label{fig:3d1}
\end{figure}

We extract comment activity log of 240 unique users consisting of 13000 comments from some of the top rated and most viewed videos on YouTube. Figures \ref{fig:spam},\ref{fig:percspam},\ref{fig:commvideo},\ref{fig:timediff} and \ref{fig:3d1} plot 119 users (for which the number of comments was greater than 5) across multiple dimensions and attributes. Figure \ref{fig:spam} reveals that there are several users with more than 20 comments having more than 50\% of the comments tagged with hasSpamHint flag. 

We observe several users with more than 60 comments and more than 70\% of them were marked as hasSpamHint. Figure \ref{fig:percspam} provides a different perspective that plots each user on the attribute of percentage spam and CRR (comment repetition and redundancy). Figure \ref{fig:percspam} clearly shows that users A, B, C and D have posted more than 30 comments (A 35, B 118, C 36 and D 30), have a CRR value of more than 0.7 (which means posting same comment multiple times) and have 80\% of the comments marked as spam by the moderator. Users on top right corner of Figures \ref{fig:spam} and \ref{fig:percspam} are potential spammers. We perform a manual inspection of such users and confirm the hypothesis to be true. Table \ref{tab:spam} shows an illustrative list of comments of some the users identified as spammers according to the proposed approach. 

\begin{table}[t]
\centering
\begin{tabular}{|p{3in}|} \hline
\small watch?v=JNVeTR9MhAo? \\ \hline
\small Check out my channel \\ \hline
\small Please watch my vids at airborn2048 \\ \hline
\small CAT almost BIT ME A FINGER OFF)))))   view on my channel )))) \\ \hline
\small password please !!!!!!!!!!!!!!!!!!!!!!!!!!!!!1 \\ \hline
\small watch?v=DGHC-AgB8Us? \\ \hline
\small TV4500Channels.blogspot.com \\ \hline
\small If you have any immense black ops videos please could you send them to bennyboy536@hotmail.com \\ \hline
\small This would give you publicity and increase your rep. For your videos to be posted on the top 5 amazing kills Thanks \\ \hline
\small PLZ SUBSCRIBE AND COMMENT TO MY CHANNEL PLZ GIVE ME A CHANCE AND HEAR MA SONGS \\ \hline 
\small CHECK OUT MY VIDS AND COMMENT \\ \hline
\small CHECK OUT OUR CHANNEL! IT IS SO FUNNY! PLEASE SUBSCRIBE! \\ \hline
\small chek out nikkayx26 \\ \hline
\small Make sure to check out my page for the  Exclusive Dance Battle of the Week!!! \\ \hline
\end{tabular}
\caption{An illustrative list of comments of some of the users identified as spammers in the experimental dataset}
\label{tab:spam}
\end{table}
Figure \ref{fig:commvideo} reveals users having high comment overlap and low video overlap (means several similar comment posting but in a single or small set of videos) as well as users having high comment overlap and high video overlap (a phenomenon wherein a user posts exactly same comments across multiple videos). Figure \ref{fig:timediff} reveals users posting a large number of comments in a small time interval (y-axis is log of the metric average time difference between comments in seconds). Users in the bottom right corner (below 3 in y value and above 20 in x value) of Figure \ref{fig:timediff} are potential spammers. Figure \ref{fig:3d1} is a plot of several users across 3 dimensions. Based on our manual inspection of the data, we derive the following rule to automatically classify a vector representing a YouTube user behavior across four dimensions (PCHF, ATDC, COMOVP and VIDOVP). All the users (number of comments > 5 as minimum threshold) satisfying the following rules were manually annotated as spammers. \\

\textbf{SPAMMER =} \underline{(PCHF > 70) OR (ATDC < 150)} \\
\underline{OR (COMOVP > 0.60) OR (VIDOVP > 0.60)} \\

A manual inspection of user profiles and comments demonstrate that comment spammers are prevalent in YouTube forums and the proposed heuristics (based on testing the presence of pre-defined spam indicators or markers in a users comment activity log) is reliable in spammer detection (refer to Table \ref{tab:spam}: a manual inspection of the comments posted by identified users clearly indicates spam). 
\section{Conclusions}
We describe a method (rule-based system) to automatically identify comment spammers in YouTube forums by mining comment activity log of users. Applying the proposed method on a sample dataset reveals that the technique is effective in identifying spammers. We hypothesize certain characteristics of comment spammers and perform an empirical study to test the proposed hypothesis. Our findings indicate that attributes such as presence of large number of exactly same comment in a single or across multiple videos, very small time intervals between subsequent comments and a large percentage of comments having spam hint flag are reliable indicators for categorizing YouTube forum spammers.

\scriptsize{
\bibliographystyle{plain}
\bibliography{USEWOD}
} 
\end{document}